\documentstyle[prb,aps,psfig,twocolumn]{revtex}

\begin{document}
\draft
\title{Control of Non-Resonant Effects in a
Nuclear Spin Quantum Computer \\ with a Large Number of Qubits}
\author{G.P. Berman$^\dag$, G.D. Doolen$^\dag$, D.I. Kamenev$^{\dag}$,
and V.I. Tsifrinovich$^\ddag$}
\address{$^\dag$ Theoretical Division and CNLS, Los Alamos National Laboratory,
Los Alamos, NM 87545}
\address{$^\ddag$ IDS Department, Polytechnic University, Six Metrotech
Center, Brooklyn, New York 11201}
\maketitle
\begin{abstract}
%\vspace{-5mm}
%
We discuss how to simulate simple quantum logic operations
with a large number of qubits. These simulations are needed for
experimental testing of scalable solid-state quantum computers. Quantum logic for remote qubits
  is simulated in a spin chain. Analytical estimates are presented for
possible correlated
  errors caused by non-resonant transitions. A range of parameters
  is given in which non-resonant effects can be minimized.
\end{abstract}

\vspace{-5mm}
\section*{Introduction}
\vspace{-2mm}
 Several proposals for scalable solid-state quantum computers have been
recently published \cite{kane,divincenzo,mrfm}. The simulation of quantum information
processing with a large number of qubits will be necessary to test these computers. We show in this paper how to implement the Control Not (CN) gate
between remote qubits
in a spin chain using a sequence of electromagnetic pulses. The errors
generated by non-resonant effects
  are estimated analytically and computed numerically.
  We give the probability of error as a function of the Rabi frequency
  and the number of qubits.

  The two-level approximation for an individual transition used in this paper for analysis of the
system
  (see also Ref. [4]) allows one to simulate the quantum
dynamics
  in a system with an enormous number ($2^L$, where $L$ is the number of qubits)
  of quantum states. This technique is based on selection of
  states generated as a result of the resonant or near-resonant
  transitions, while other transitions are neglected.

  In Sec. I we give the
  Hamiltonian of the nuclear spin quantum computer.
  In Sec. II we use the resonance approximation to show that each
quantum state
  interacts resonantly only with a single state. This allows us to
  decrease from exponential in $L$ to polynomial in $L$ the number of differential equations in the system.
  In Sec. III we use the solution of the problem in the resonance
  approximation to estimate error probabilities in the implementation of a CN gate
  caused by finite detuning from exact resonance.

\vspace{-5mm}
\section{Dynamics of the spin chain}
\vspace{-2mm}
We consider a chain of identical nuclear spins placed in a nonuniform  high external magnetic field $B$, with a uniform gradient. 
The nuclear magnetic resonance (NMR) frequency for the $k$th spin is $\omega_k=\gamma_n B_k$, where
$\gamma_n$ is the nuclear gyromagnetic ratio and $B_k$ is the $z$-component of the
magnetic field at the location of the $k$th spin.
The gradient of the magnetic field (in the direction of the chain) 
provides a shift of $\omega_k$ by the value $\delta\omega$
between the neighboring spins. (For the physical parameters see Ref.~[4].)

The Hamiltonian of the spin chain in an external radio-frequency (rf)
field is,
\vspace{-3mm}
\begin{equation}
\label{H}
H_n=-\sum_{k=0}^{L-1}\omega_kI_k^z-2J\sum_{k=0}^{L-1}I_k^zI_{k+1}^z+V_n=
H_0+V_n,
\end{equation}
where $J$ is the Ising interaction constant and $I_k^z$ is the operator of the
$z$-component of spin $1/2$. The operator, $V_n$, describing the interaction
of spins with the {\it rf} pulses, can be written in the form, \cite{book}
$
\label{V}
V_n=-(\Omega_n/2)
\sum_{k=0}^{L-1}[I_k^-\exp\left(-i(\nu_n t+\varphi_n\right)+
I_k^+\exp(i\nu_n t+$\\
$\varphi_n)],$
where $\Omega_n$, $\nu_n$ and $\varphi_n$ are the Rabi frequency,
the frequency and the phase of the $n$th pulse;
$I_k^\pm=I_k^x\pm I_k^y$. 
Each quantum state of the spin chain can be described as a superposition
of eigenstates of $H_0$, for example,
$
|00\dots 00\rangle,\,|00\dots 01\rangle,
$
and so on, where the state $|\dots 0_k \dots\rangle$
corresponds to the direction of
the $k$th nuclear spin along the direction of the
magnetic field (spin up), and the state
$|\dots 1_k \dots\rangle$ corresponds to the spin being in the opposite direction (spin down).
The wave function, $\Psi$, of the spin chain can be written in the interaction representation as a linear combination of the
individual states,
$
\Psi(t)=\sum_pC_p(t)|p\rangle\exp(-iE_pt),
$
where $E_p$ is the energy of the state $|p\rangle$. The contribution of
each quantum state to the wave function, $\Psi(t)$, is given
by the coefficient $C_p(t)$. The quantity $|C_p(t)|^2$ is the probability
of finding the spin chain in the state $|p\rangle$ at time $t$.

The Schr\"odinger
equation for the coefficients $C_p(t)$ has the form (we put $\varphi_n=0$),
$
i\dot C_p(t)=\sum_{m=0}^{2^L-1}V^n_{pm}
\exp[i(E_p-E_m)t+ir_{pm}\nu_nt]C_m(t),
$
where $r_{pm}=\mp 1$ for $E_p>E_m$ and $E_p<E_m$, respectively.
$V^n_{pm}=-\Omega_n/2$ for the states $|p\rangle$ and $|m\rangle$
connected by a single-spin transition. $V^n_{pm}=0$ for all other
states.
\vspace{-5mm}
\section{The approximate solution}
\vspace{-2mm}
The Schr\"odinger equation can be numerically integrated only
for a spin chain with small enough number of spins since the number
of states increases exponentially with $L$.
The problem can be simplified when the Rabi frequency, $\Omega_n$,
is much less than the difference, $\delta\omega$, between the
NMR frequencies of the neighboring spins, $\Omega_n\ll \delta\omega$.
Suppose that a pulse is resonant with the $k$th spin in the chain,
$\nu_n\approx \omega_k$.
Then one can admit that this pulse affects in some approximation only the $k$th spin
in the chain and does not interact with other spins.
In this case, only the term with the energy
$E_{m'}\approx E_p+\nu_n$ effectively contributes
to the right-hand side of the Schr\"odinger equation,
so that the states $|m'\rangle$ and $|p\rangle$ are
connected by a single-spin flip of the $k$th spin.
In this case, we reduce the Schr\"odinger equation
to the set of two differential equations,
\vspace{-3mm}
$$
i\dot C_m(t)=-{\Omega_n\over 2}
\exp[i(E_p-E_m-\nu_nt]C_m(t),
$$
\begin{equation}
\label{Sch1}
i\dot C_p(t)=-{\Omega_n\over 2}
\exp[-i(E_p-E_m-\nu_nt]C_p(t),
\end{equation}
%\vspace{-3mm}
where $E_p>E_m$, and $|p\rangle$ and $|m\rangle$ are two stationary states which
are connected by a single-spin transition of the $k$th spin,
$\nu_n\approx \omega_k$.

 The solution of Eq.~(\ref{Sch1}), for initial conditions, $C_m(t_{n-1})=1,\quad C_p(t_{n-1})=0$, after the action of the $n$th pulse has the form,
\vspace{-2mm}
$$
C_m(t_n)=\left[\cos\left(\lambda_n\tau_n/ 2\right)+
i{\Delta_n\over\lambda_n}\sin\left(\lambda_n\tau_n/2\right)\right]
e^{\left({-i\tau_n\Delta_n\over 2}\right)},
$$
\begin{equation}
\label{map}
C_p(t_n)=i{\Omega_n\over\lambda_n}\sin\left(\lambda_n\tau_n/2\right)
e^{\left(it_{n-1}\Delta_n+i{\tau_n\Delta_n\over 2}\right)},
\end{equation}
where $t_{n-1}$ and $t_{n}$ are the beginning and the and of the $n$th pulse,  $\tau_n$ is the duration of the $n$th pulse, $\Delta_n=E_p-E_m-\nu_n$ (we do not indicate the dependence of $\Delta_n$ on the indices $p$ and $m$), $\lambda_n=\sqrt{\Omega_n^2+\Delta_n^2}$ is the
precession frequency in the frame rotating with the frequency $\nu_n$. Suppose that at $t_{n-1}$ the system was in the state $|m\rangle$. In the case of the exact resonance
($\Delta_n=0$) and for $\Omega_n \tau_n=\pi$ (a $\pi$ pulse) Eqs.~(\ref{map})
describe the resonant transition from the state $|m\rangle$ to the state
$|p\rangle$. If the frequency of the next pulse also satisfies the
resonance condition, $\Delta_{n+1}=E_l-E_p-\nu_{n+1}=0$, then the system
with probability equal to unity will transform to the state $|l\rangle$, and so on.

\vspace{-5mm}
\section{Errors in creation of entangled state for remote qubits}
\vspace{-2mm}
The advantages given by the two-level approximation described above
are obvious:
(a) we solve the problem by using the discrete map (\ref{map}) instead of
integrating differential equations; (b) we can
consider only the states with large enough probabilities and
neglect all other states in a controlled way; (c) we can estimate errors caused by non-resonant effects and minimize them by choosing optimal
parameters.

In this paper we will estimate the error in implementation
of a unitary operation for remote qubits (which is a particular case of the well-known Control-Not (CN) gate) using a sequence of $\pi$-pulses.
This CN gate is defined as the unitary operator, $U$, with the
following properties:
\begin{mathletters}
\begin{equation}
\label{U1}
U|0_{L-1}0_{L-2}\dots 0_{1}0_0\rangle=e^{i\varphi_1}|0_{L-1}0_{L-2}\dots 0_10_0\rangle,
\end{equation}
\vspace{-8mm}
\noindent
\begin{equation}
\label{U2}
U|1_{L-1}0_{L-2}\dots 0_{1}0_0\rangle=e^{i\varphi_2}|1_{L-1}0_{L-2}\dots 0_11_0\rangle,
\end{equation}
\end{mathletters}
where $\varphi_{1}$ and $\varphi_{2}$ are known phases.
The target qubit ($0$) should change its state only when the control qubit ($L-1$) is in the
state $|1\rangle$. The operator, $U$, can be used to create an entangled state for remote qubits: $U(\alpha|0_{L-1}...0_0\rangle+\beta|1_{L-1}...0_0\rangle)=\alpha e^{\varphi_1}|0_{L-1}...0_0\rangle+\beta e^{\varphi_2}|1_{L-1}...1_0\rangle$.
To accomplish the (4b) operation in the system described by the 
Hamiltonian (1), we choose a sequence of $\pi$-pulses with resonant frequencies (for which all $\Delta_n=0$).
In this case, the state $|10\dots 00\rangle$ transforms to the state
$|10\dots 01\rangle$, with probability equal to unity, by the following scheme:
$
|1000\dots 0\rangle\rightarrow|1100\dots 0\rangle\rightarrow
|1110\dots 0\rangle\rightarrow$
$
|1010\dots 0\rangle\rightarrow|1011\dots 0\rangle\rightarrow
|1001\dots 0\rangle\rightarrow
$
$
\dots\rightarrow|100\dots 11\rangle\rightarrow|100\dots 01\rangle.
$
The sequence of pulses which realizes this protocol has the following
form: $\nu_1=\omega_{2L-2}$, $\nu_2=\omega_{2L-3}$, $\nu_3=\omega_{2L-2}-2J$,
$\nu_{2L-4}=\omega_6$,
\dots, $\nu_{2L-4}=\omega_0-J$, $\nu_{2L-3}=\omega_1$.
If we apply the same protocol to the state $|00\dots 00\rangle$, then
with some probability the system will remain in this state because all transitions are non-resonant with the detuning
$\Delta_n=\pm J,\pm 2J$. Since $\Delta_n\ne 0$ these
non-resonant transitions have the probabilities (see Eq.~(\ref{map})),
\begin{equation}
\label{epsilon}
\varepsilon_n=(\Omega_n/\lambda_n)^2\sin^2(\lambda_n \tau_n/2).
\end{equation}

In this paper we study the errors generated during
the operation (\ref{U1}) under the condition that the transitions in Eq.~(\ref{U2}) are resonant.
The operator $U$ in Eqs. (\ref{U1}) and (\ref{U2}) can be written as a product,
$U=U_{2L-3}U_{2L-4}\dots U_{2}U_1$, of operators of individual pulses $U_n$, where
$n=1,\,,2,\dots,2L-3$. We take the Rabi frequency to be the same,
$\Omega_n=\Omega$, for all pulses. Then the error, $\varepsilon_n$,
generated by the $n$th pulse is defined only by the detuning, $\Delta_n$.
The values of detuning, $|\Delta_n|$, are the same for all pulses,
$|\Delta_n|=J$, except for the third pulse, where $|\Delta_3|=2J$.
We denote $\varepsilon_n\equiv\varepsilon$ for $n\ne 3$ and
$\varepsilon_{n=3}\equiv\varepsilon'$.

Each unwanted state generated by the $n$th pulse can produce
other unwanted states under the action of the other $(n+m)$ pulses.
This causes the generation of a  hierarchy of different states.
 Instead of one ground state, which should be the result of (4a), we generate a series of different unwanted states.

\vspace{-2mm}
\begin{figure}
\centerline{\psfig{file=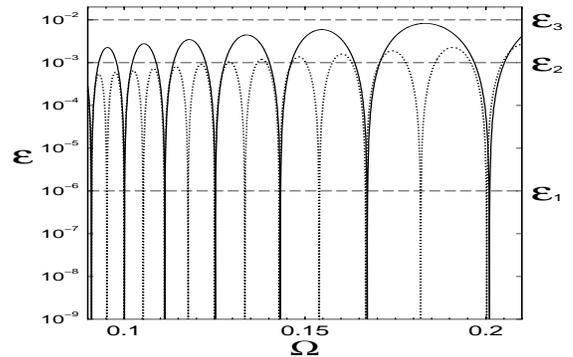,width=7.5cm,height=5cm}}
%\vspace{-5mm}
\caption{Dependence of $\varepsilon$ (solid line) and $\varepsilon'$ (dotted line) on
$\Omega$. $J=1$.}
\label{fig:1}
\end{figure}
\vspace{-2mm}

Since the
dynamics is generated by the discrete map (\ref{map}), probability of
error can be estimated analytically. Suppose that we can measure only 
the states with probability $P\ge P_0$, where $P_0$ can be defined by the experimental conditions. Then, if
$\varepsilon,\,\varepsilon'< \varepsilon_1=P_0$, we shall not find in the system
any unwanted states and the CN gate
is realized with probability equal to unity. In Fig.~1, we plot
$\varepsilon$ (solid line) and $\varepsilon'$ (dotted line) as  functions of
$\Omega$. We suppose that $\varepsilon_1=P_0=10^{-6}$. One can see from Fig.~1 that $\varepsilon(\Omega)$ and $\varepsilon'(\Omega)$
are less than $\varepsilon_1$ only in very narrow regions of the values
of $\Omega$ (in the vicinity of $\Omega_k=|\Delta_n|/\sqrt{4k^2-1}$, $k=1,2,...$ \cite{1,book}). Hence, high precision is required to implement
the CN gate, (\ref{U1}), (\ref{U2}), without error.

Let us consider now in detail the case of arbitrary $\varepsilon$: $0\le \varepsilon\le 1$, and explain how
the errors can be
estimated analytically in the operation (\ref{U1}). The first
pulse $U_1$ generates one
unwanted state $U_1|0000\dots 0\rangle\rightarrow|0100\dots 0\rangle$
with the probability $\varepsilon$. The probability to remain in the ground state
is $1-\varepsilon$. The second pulse, $U_2$, generates an additional unwanted
state from the ground state,
$U_2|0000\dots 0\rangle\rightarrow|0010\dots 0\rangle$, with probability
$\varepsilon(1-\varepsilon)$, and transforms the state
$|0100\dots 0\rangle$ to the state $|0110\dots 0\rangle$ without
generating additional states, since the second pulse is resonant to this transition. The probability of remaining in the ground state
is now $(1-\varepsilon)^2$.
The total number of states after the action of two
pulses is three.
The probabilities of all generated
states after the action of the $U_3$ pulse are:
\begin{tabbing}
\label{tab}
$|0000\dots 0\rangle$: \hspace{1cm}\= $(1-\varepsilon)^2(1-\varepsilon')$, \\
\hspace{1mm}$|0100\dots 0\rangle$: \> $\varepsilon'(1-\varepsilon)^2$, \\
\hspace{1mm}$|0010\dots 0\rangle$: \> $\varepsilon(1-\varepsilon+\varepsilon^2)$, \\
\hspace{1mm}$|0110\dots 0\rangle$: \> $\varepsilon(1-\varepsilon^2)$.
\end{tabbing}
(Note, that above expressions can be considered only as estimates because we estimated probabilities instead of complex amplitudes.)
One can continue to calculate the states generated by subsequent pulses
and estimate their probabilities in
a similar way. However, for $\varepsilon,\,\varepsilon'\ll 1$, one can
omit the contribution from the higher order terms in $\varepsilon$, and neglect the states with the probabilities less than $P_0$.

Suppose, that $\varepsilon'\approx\varepsilon$ and
$\varepsilon_1<\varepsilon<\varepsilon_2\equiv \sqrt{P_0}$. In this case, each pulse generates
only one unwanted state from the ground state. Then, only
\vspace{-2mm}
\begin{equation}
\label{N1}
N_1(L)=2L-3
\end{equation}
%\vspace{-1mm}
unwanted states will be generated with probabilities
(with the accuracy up to $\varepsilon^2$)
$\varepsilon$, $\varepsilon(1-\varepsilon)$, $\varepsilon(1-2\varepsilon)$, \dots,
$\varepsilon(1-(2L-4)\varepsilon)$. After the action of $2L-3$ pulses there are
$L-1$ unwanted states of the form,
\begin{equation}
\label{1}
|011\dots 11\rangle,~|0011\dots 11\rangle ,...,
|00\dots 011\rangle,~ |00\dots 01\rangle,
\end{equation}
with the target spin in the state $|1\rangle$, and $L-2$ unwanted states of the form,
\begin{equation}
\label{0}
|011\dots 110\rangle,~|0011\dots 110\rangle,~\dots,~
|00\dots 010\rangle,
\end{equation}
with the target spin in the state $|0\rangle$. As one can see from Eqs. (\ref{1}) and (\ref{0}),  already in the first order of $\varepsilon$ many states have correlated error with many qubits in excited states. The estimate of the total probability of unwanted states is
\vspace{-3mm}
\begin{equation}
\label{epsilon}
P_1({\cal E})=\varepsilon\sum_{n=0}^{2L-4}(1-n\varepsilon)\approx
{\cal E}\left(1-\frac 12{\cal E}\right),
\end{equation}
where ${\cal E}=(2L-3)\varepsilon$.

The probability of unwanted effect of finding a target qubit in the state $|1\rangle$ while
the control qubit is in the state $|0\rangle$
(i.e. the probability of the process
$U|0\dots 0\rangle\rightarrow|0\dots 1\rangle$) is, 

\vspace{-4mm}
\begin{equation}
\label{epsilon1}
{\cal P}_1({\varepsilon})=\varepsilon+\varepsilon\sum_{n=0}^{L-3}(1-(2n+1)\varepsilon)
\approx\Gamma(1-\Gamma),
\end{equation}
%\vspace{-1mm}
where $\Gamma=(L-2)\varepsilon$.

\vspace{-3mm}
\begin{figure}
\mbox{\hspace{-4mm}\psfig{file=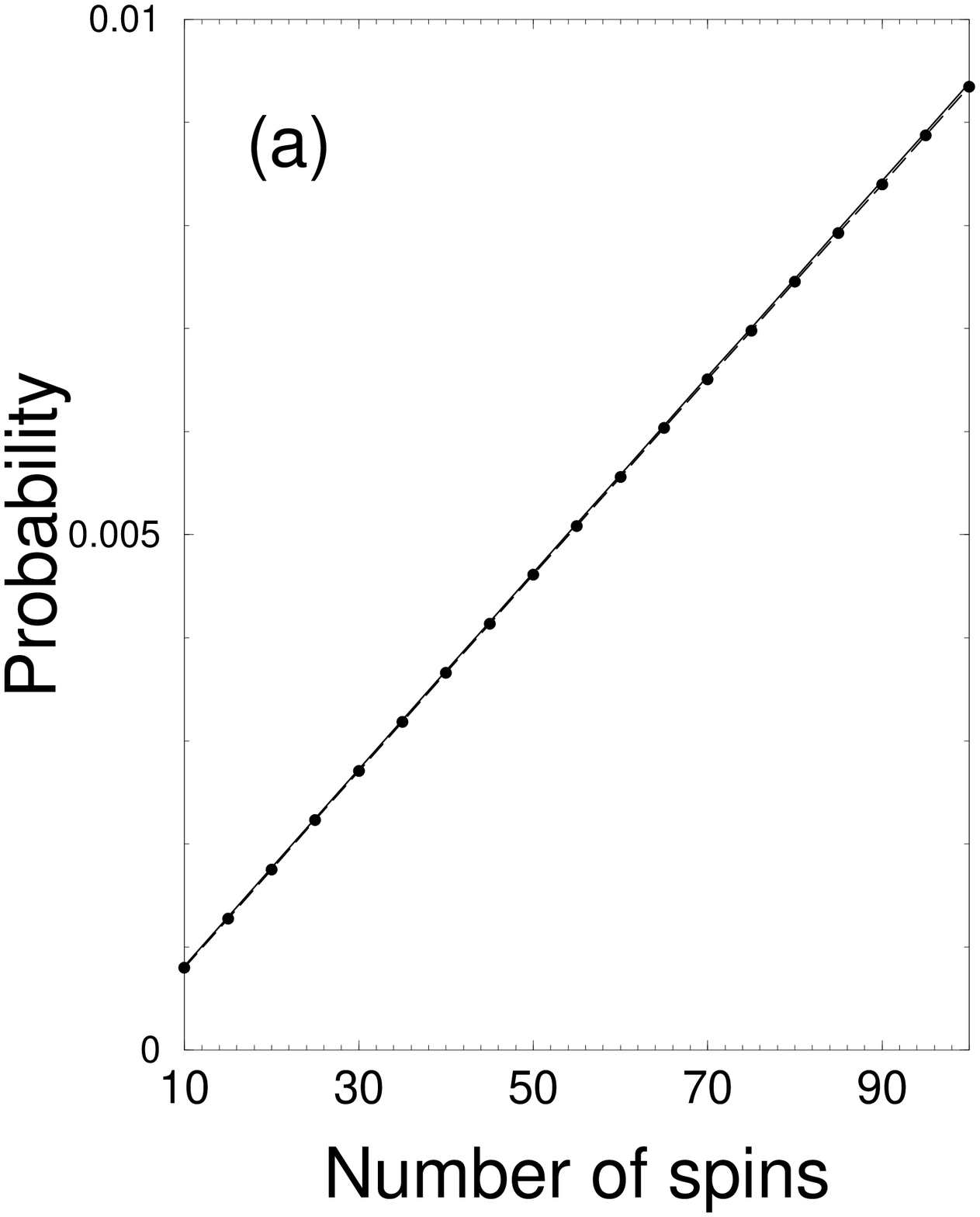,width=4.2cm,height=5cm}
\hspace{2mm}\psfig{file=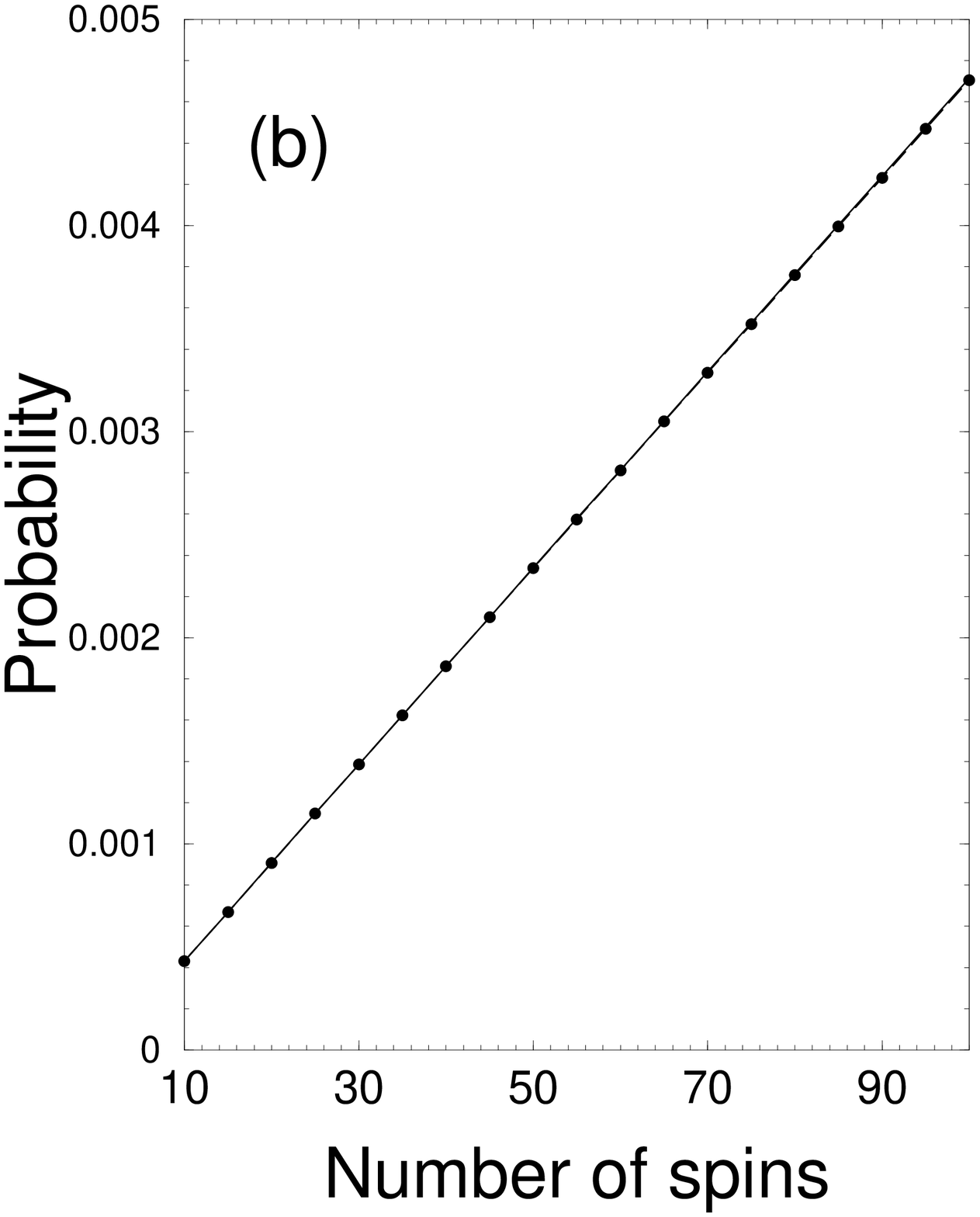,width=4.2cm,height=5cm}}
\vspace{-2mm}
\caption{(a): The total probability, $P_1$, of unwanted states in the execution of the CN logic gate as a function of the number of qubits, $L$, in the spin chain.
Solid line: analytical estimate (\ref{epsilon}), filled circles connected by dashed line: numerical results.
(b) Probability, ${\cal P}_1$, of observing the target spin in the state $|1\rangle$,
while the control qubit is in the state $|0\rangle$, after the CN operation; solid line: analytical estimate (\ref{epsilon1}), filled circles connected by dashed line: numerical results. $J=1$, $\Omega=0.0906$.}
\label{fig:2}
\end{figure}

In Figs.~2~(a),~(b) we plot the probabilities, $P_1$ and ${\cal P}_1$, as 
functions of the number of spins, $L$, in the chain, for
$\Omega=0.0906$, so that
$\varepsilon_1<\varepsilon(\Omega)=4.78\times 10^{-5}<\varepsilon_2=10^{-3}$
 ($\varepsilon'(\Omega)=3.23\times 10^{-5}$). The number of unwanted
states in our calculations was exactly equal to $N_1$ in Eq.~(\ref{N1}).
From Figs.~2~(a) and (b) one can see
that the analytical estimates agree with the results of numerical calculations.

In a similar way one can calculate the probabilities of unwanted states
for the case $\varepsilon_2<\varepsilon<\varepsilon_3\equiv (P_0)^{1/3}$. In this
situation one should take into account states with the probabilities
up to $\varepsilon^2$. Each of these states is generated as a result of two
successive non-resonant processes. However,
since the ground state generates only unwanted states with
the probabilities of order $\varepsilon$, the total probability
of unwanted states will be given again by Eq.~(\ref{epsilon}).
But the number of generated states increases and at the condition, $\varepsilon_2<\varepsilon<\varepsilon_3$, it is proportional
to $L^2$.

%The total number of such states is,
%\begin{equation}
%\label{N2}
%N_2(L)=(L-3)(L-1)/2-2,~L\ge 5,
%\end{equation}
%where $[\dots]$ indicates the integer part.

In Fig.~3 we compare the numerical results with
the analytical estimates for the total probability of unwanted states
given by Eq.~(\ref{epsilon}) when the value $\varepsilon$ is
relatively large.
From Fig.~3 one can see that formula (\ref{epsilon})
correctly describes the behavior of the system for large
value of $\varepsilon$. When $L$ becomes large enough, one should also
include into consideration in Eq. (\ref{epsilon}) the terms of order
$(L\varepsilon)^3$.

\begin{figure}
\vspace{-3mm}
%\hspace{-4mm}
\centerline{\psfig{file=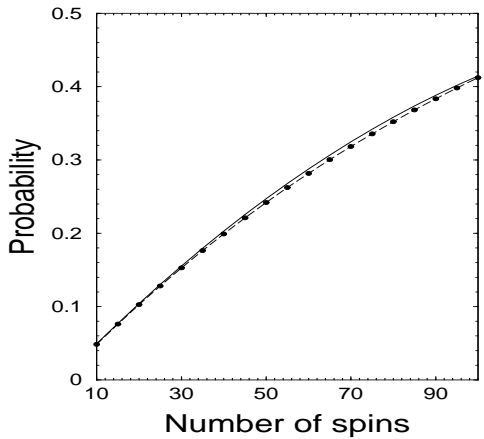,width=6.2cm,height=6cm}}
%\hspace{2mm}\psfig{file=eeb.ps,width=4.2cm,height=4.8cm}}
%\vspace{-2mm}
\caption{The total probability of unwanted states for
the case of relatively large $\varepsilon$, when
$\varepsilon_1<\varepsilon<\varepsilon_2$.
Solid line:
analytical estimate (\ref{epsilon}),
filled circles connected by dashed line: numerical results.
$J=1$, $\Omega=0.20844$,
$\varepsilon(\Omega)=2.98\times10^{-3}$,
$\varepsilon'(\Omega)=2.41\times10^{-3}$.}
\label{fig:3}
\end{figure}

The probabilities of unwanted states are shown in Figs.~4~(a) and (b) in two different scales. Two
types of states are clearly seen. The states in Fig. 4 (a) are generated 
as a result of one non-resonant transition, so the probabilities of these states are proportional to $\varepsilon$. On the other hand, the states in Fig. 4 (b) are generated by two successive non-resonant transitions, and they have the probabilities of the order $\varepsilon^2$.

\begin{figure}
\vspace{-3mm}
\mbox{\hspace{-4mm}\psfig{file=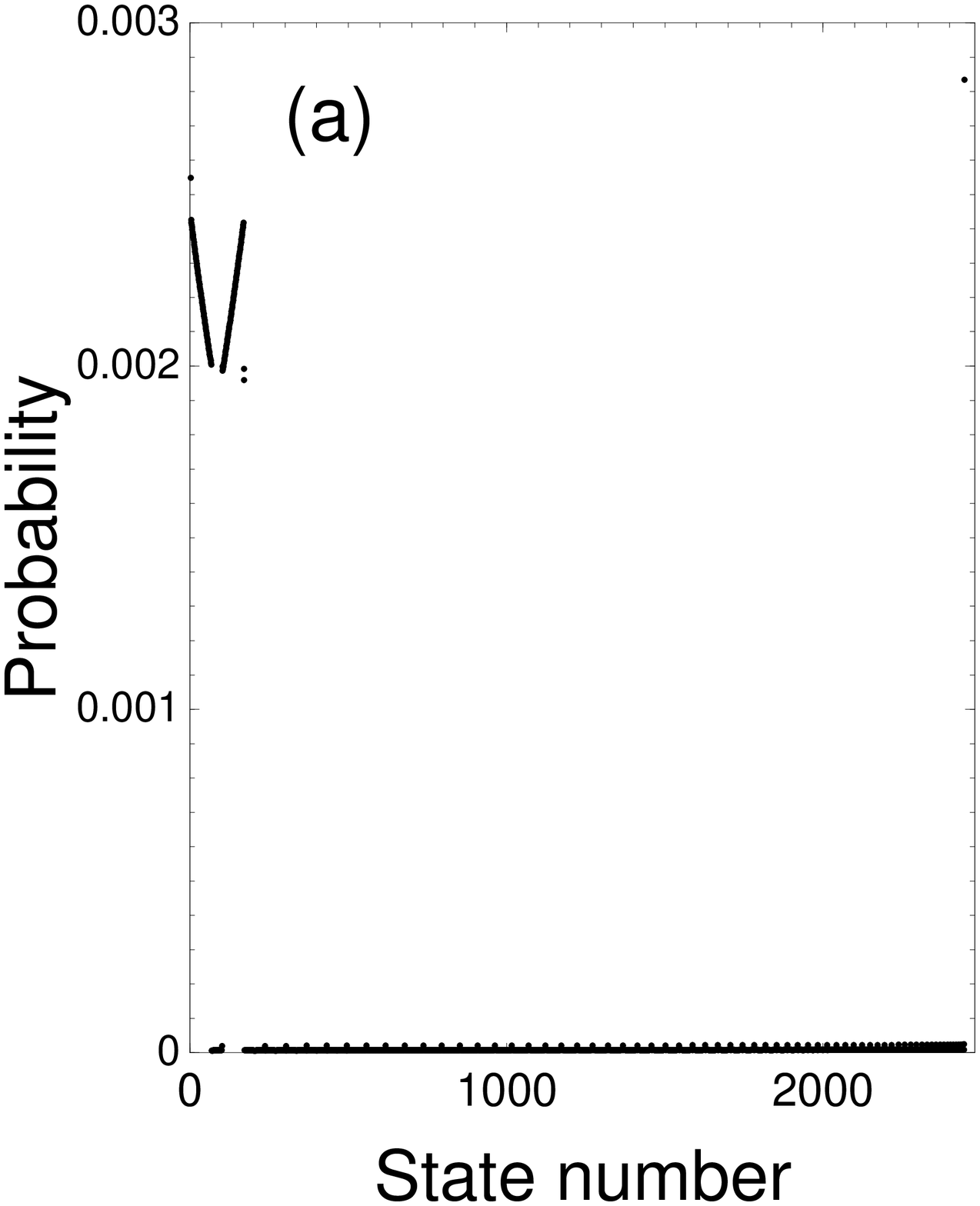,width=4.2cm,height=5.2cm}
\psfig{file=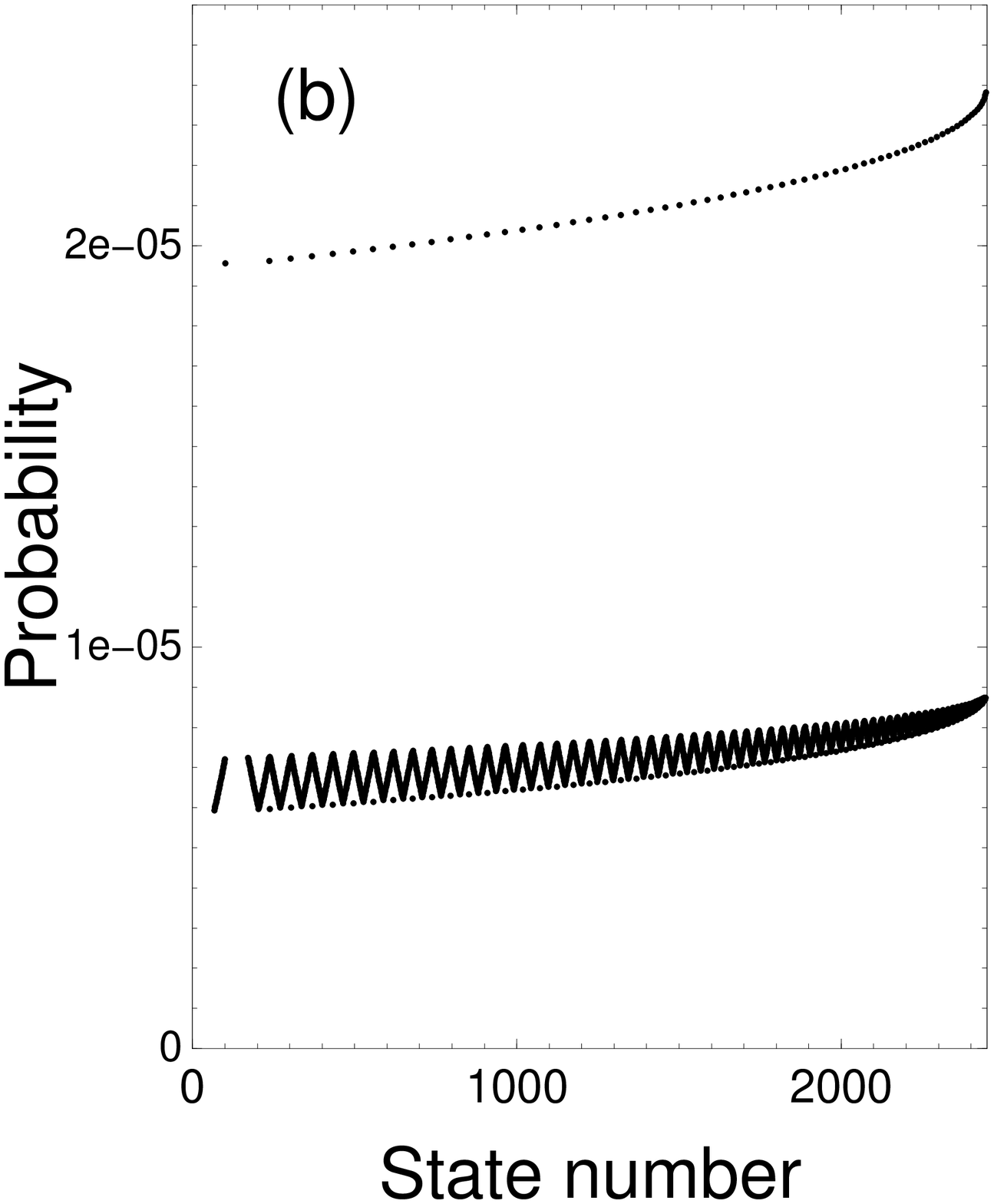,width=4.2cm,height=5cm}}
\vspace{2mm}
\caption{The probabilities of generated states in two different scales.
$L=70$. All other parameters are the same as in Fig. 3.}
\label{fig:4}
\end{figure}

\vspace{-5mm}
\section{Conclusion}
\vspace{-2mm}
We have demonstrated how one can simulate a simple logic operation for a large number of qubits. 
We analyzed the probability of correlated errors in the implementation of
  a CN gate between remote qubits in a spin chain with a large number of
qubits.
  In the limit when the Rabi frequency $\Omega$ is much less than the
difference
  between the NMR frequencies of neighboring spins, it
  is shown that the probability of error, $P$, is mainly defined by the
  small parameter $\varepsilon$ defined by (5) and by the number of qubits in the spin
chain.
  It is demonstrated that at definite values of $\Omega$ the value of $\varepsilon$
  is small and there is no correlated error in the implementation of the CN gate. The
total
  probability of unwanted states at other values of $\varepsilon$ is
estimated
  analytically and computed numerically.

  Since the Hamiltonian of the system only allows
  transitions between the states connected by a single spin
transformation,
  and because the electromagnetic field interacts resonantly only with
  one spin in the spin chain, the probability of error cannot
accumulate
  in a definite single unwanted state. The probability of each
unwanted state
  is always less than the value of the small parameter of the
problem
  $\varepsilon(\Omega)$. The considered approach can be useful for
experimental testing of scalable solid-state quantum computers.

The paper was supported by the Department of Energy (DOE) under
contract W-7405-ENG-36 and by the National Security Agency (NSA), and by the Advanced Research and Development Activity (ARDA).

\end{document}